\documentclass[prd,12pt,tightenlines,raggedbottom]{revtex4-2}

\usepackage[usenames,dvipsnames]{xcolor}
\usepackage{hyperref}
\hypersetup{
	colorlinks=true,
	urlcolor=Maroon,
	linkcolor=RoyalBlue,
	citecolor=Maroon,
	bookmarks=true,
	pdftitle={Bounds on Crossing Symmetry},
	pdfauthor={Sebastian Mizera},
	pdfdisplaydoctitle=true,
	pdfstartview=FitH
}

\usepackage{amsmath}
\usepackage{amsfonts}
\usepackage{amssymb}
\usepackage{mathtools}
\usepackage{microtype}
\usepackage{bm}
\usepackage[utf8]{inputenc}
\usepackage{blkarray}
\usepackage{bigstrut}
\usepackage{nccmath}
\usepackage{enumitem}
\usepackage{braket}
\usepackage{dsfont}
\usepackage[export]{adjustbox}
\usepackage[english]{babel}
\usepackage{physics}
\usepackage{scalerel}
\usepackage[font=small,labelfont=bf,labelsep=period]{caption}
\usepackage[margin=1in]{geometry}

\usepackage[nodisplayskipstretch]{setspace}
\allowdisplaybreaks
\graphicspath{{figures/}}

\def \be {\begin{equation}}
\def \ee {\end{equation}}
\def \nn {\nonumber}

\def \D {\mathrm{D}}
\def \E {\mathrm{E}}
\def \L {\mathrm{L}}
\def \U {\mathcal{U}}
\def \F {\mathcal{F}}
\def \d {\mathrm{d}}
\def \eps {\varepsilon}
\def \V {\mathcal{V}}
\def \G {\mathcal{G}}
\def \zero {\scaleto{(0)}{7pt}}
\def \one {\scaleto{(1)}{7pt}}
\def \tast {\scaleto{(\ast)}{7pt}}
\def \talpha {\scaleto{(\alpha)}{7pt}}

\begin{document}

\title{\Large Bounds on Crossing Symmetry}
\author{Sebastian Mizera\\ \footnotesize\href{mailto:smizera@ias.edu}{\texttt{smizera@ias.edu}}}
\affiliation{Institute for Advanced Study, Einstein Drive, Princeton, NJ 08540, USA \vspace{1em}}

\begin{abstract}\linespread{1}\selectfont
Proposed in 1954 by Gell-Mann, Goldberger, and Thirring, crossing symmetry postulates that particles are indistinguishable from anti-particles traveling back in time. Its elusive proof amounts to demonstrating that scattering matrices in different crossing channels are boundary values of the same analytic function, as a consequence of physical axioms such as causality, locality, or unitarity. In this work we report on the progress in proving crossing symmetry on-shell within the framework of perturbative quantum field theory. We derive bounds on internal masses above which scattering amplitudes are crossing-symmetric to all loop orders. They are valid for four- and five-point processes, or to all multiplicity if one allows deformations of momenta into higher dimensions at intermediate steps.
\end{abstract}

\maketitle
\pagestyle{plain} 

\section{Introduction}

Ever since its introduction in 1954 by Gell-Mann, Goldberger, and Thirring, crossing symmetry has been widely believed to be a fundamental property of Nature  \cite{GellMann:1954db,GellMann:1954kc,GellMannWebOfStories}. It postulates that particles are equivalent to anti-particles with opposite energies and momenta, or---more precisely---that their scattering amplitudes can be analytically continued between different crossing channels. It is routinely taken as an assumption in various bootstrap approaches to the scattering matrix theory, see, e.g., \cite{Arkinson:1968hza,Martin:1969ina,Adams:2006sv,Paulos:2017fhb,Guerrieri:2018uew,Bellazzini:2020cot,Tolley:2020gtv,Guerrieri:2020bto,Caron-Huot:2020cmc,Tourkine:2021fqh}. Yet, crossing symmetry does not \emph{directly} follow from any physical principle and there is only a limited amount of theoretical evidence that it holds in the Standard Model or even a generic quantum field theory. For instance, crossing between $2\to3$ and $3\to2$ processes has never been proven, and neither has any case involving massless particles.

The extent to which we can be certain that crossing symmetry is true non-perturbatively stems from the work of Bros, Epstein, and Glaser, who proved it in the case of $2\to2$
\cite{Bros:1964iho,Bros:1965kbd}
and $2\to3$ \cite{Bros:1972jh,Bros:1985gy} scattering in local quantum field theories with a mass gap, hinging on the assumptions of causality and unitary. Since within the Lehmann--Symanzik--Zimmerman formalism scattering amplitudes do not converge on-shell, one is forced to consider off-shell Green's functions, which \emph{can} be defined in a certain region of the complexified momentum space \cite{Steinmann1960a,Steinmann1960b,ruelle1961connection,doi:10.1063/1.1703695,araki1960properties}. We briefly review this point in App.~\ref{app:conjecture}. At this stage one is tasked with a purely geometric problem of showing that the envelope of holomorphy of this domain intersects physical regions in all crossing channels on the correct sheet, e.g., using versions of the edge-of-the-wedge theorem. The connection between scattering amplitudes in different channels is achieved via a complex kinematic region of large center-of-mass energy. For reviews see \cite{Epstein:1966yea,Sommer:1970mr,bogolubov1989general} and \footnote{See also \cite{10.1007/3-540-09964-6_319} for progress using the assumption of asymptotic completeness and \cite{Eden:1966dnq,Iagolnitzer:1978qv} for reviews of previous incomplete attempts at proving crossing symmetry.}. Such proofs are prohibitively long and technical \cite{Bros:1964iho,Bros:1965kbd,Bros:1972jh,Bros:1985gy}, and while in principle there is no obstruction to attempting generalizations to higher-point cases, they would certainly not improve our physical understanding of crossing symmetry.

In view of these difficulties, Witten proposed to prove crossing symmetry entirely \emph{on-shell} in perturbation theory, where one might reasonably hope for a simpler and more physical derivation that could potentially extend to higher multiplicity. While work on this problem is ongoing and will be published elsewhere, the purpose of this letter is to demonstrate that even using simple arguments one can put ${\cal O}(1)$ bounds on the ratios of masses above which crossing symmetry is satisfied to all loop orders.

Since for a CPT-invariant theory crossing is already apparent on the level of Feynman diagrams, the challenge lies in showing that Feynman \emph{integrals} cannot develop singularities when continued between any pair of crossed processes. To make the problem well-defined we assume that any overall divergences (such as infrared or ultraviolet), if present, have been regularized or renormalized. As a consequence, one has to consider scalar diagrams of all graph topologies with an arbitrary number of loops and external legs $n$. To each of them we can assign the function
\be\label{eq:intro-V}
\V = \sum_e \alpha_e (q_e^2 - m_e^2),
\ee
which can be thought of as the localized worldline action. Here $\alpha_e$ are the Schwinger proper times associated to the internal edges $e$, while $q_e^\mu$ and $m_e$ are the momenta and masses flowing through them, sourced by the external momenta $p_i^\mu$. Due to homogeneity in $\alpha_e$'s, extremizing the action requires $\V=0$, which is a \emph{necessary} condition for a singularity, equivalent to putting propagators on-shell.

Were it not for the requirement of causality, scattering amplitudes would be analytic along complex paths connecting \emph{any} two real non-singular points in the space of kinematic invariants $p_i {\cdot} p_j$, because along such a deformation
\be
\sum_e \alpha_e \big|q_e^2 - m_e^2\big|^2 > 0,
\ee
ergo, it is impossible to simultaneously put all propagators on-shell. However, such analytic continuations generically violate causality, which requires that $\Im \V > 0$ when approaching physical points, as dictated by the $i\eps$ prescription. Its consistent implementation is what puts bounds on crossing symmetry.

One way of ensuring causality is analytic continuation via a region where $\V < 0$ for all values of Schwinger parameters. We will show it implies that the internal masses $m_e$ cannot be too light, or more precisely
\be\label{eq:intro-m-constraint}
m_e \,>\, \frac{\sqrt{n}}{2\sqrt{2}} \sqrt{\max_i \left( M_i^2,\; \tfrac{{\sum}_{j}M_j^2 - 2M_i^2}{n-2}\right)},
\ee
where $M_i$ are the external masses.
These are bounds for crossing symmetry to be satisfied on-shell to all loop orders in perturbation theory.

For instance, the above bounds are satisfied for scattering of massless particles with all the exchanged states having arbitrary non-zero mass. This result implies crossing symmetry for a range of low-energy effective field theories, which at present does not have a counterpart on the non-perturbative level \cite{Bros:1965kbd,Bros:1985gy}.

For $n=4,5$ in the equal mass case, $M_i = M$, we have respectively
\be\label{eq:intro-bound-45}
m_e \,\gtrsim\, 0.707 M, \qquad m_e \,\gtrsim\, 0.791 M,
\ee
and for scattering of the lightest states, i.e., $m_e \geqslant M$, crossing symmetry is valid for $n<8$.
A closely related kinematic region was previously investigated in the context of dispersion relations, majorization of Feynman diagrams, and related topics for $n=4$ with the same bound in the equal-mass case \cite{Wu:1961zz}, and for $n=5$  \cite{Wu:1961zz,doi:10.1063/1.1703779,PhysRev.132.902,boyling1964hermitian} without attempts to put bounds.

The above strategy relies on linear deformations of the kinematic invariants $p_i {\cdot} p_j$ rather than the momenta $p_i^\mu$ themselves. The advantage of doing so is that we can continue between crossed processes involving a different number of incoming/outgoing particles, such as a continuation from $2\to3$ to $3\to2$ scattering. A disadvantage of this approach lies in the fact that along the deformation the momenta $p_i^\mu$ will in general span an $(n{-}1)$-dimensional space. Hence for $n \geqslant 6$ the proof requires deformations of momenta into higher dimensions at intermediate steps. While it certainly makes sense when we treat scattering amplitudes as functions of complex variables, the physical interpretation is obscured.

We work in Minkowski space with mostly-minus signature and use conventions where incoming momenta are denoted with $p_i^\mu$ and outgoing $-p_i^\mu$, such that the momentum conservation reads $\sum_i p_i^\mu = 0$.

\section{Review of Feynman Integrals and Their Singularities}

We find it most intuitive to interpret Feynman diagrams in the worldline formalism, where Schwinger proper times $\alpha_e$ are the only dynamical variables. A scalar diagram with $n \geqslant 4$ external legs, $\E$ internal edges (propagators), and $\L$ loops in $\D$ space-time dimensions can be written as
\be\label{eq:exponential}
\int_{0}^{\infty} \frac{\d^{\E}\alpha_e}{\U^{\D/2}}\, e^{i\V /\hbar},
\ee
where $\V$ is the localized action and $\U$ is the determinant of the Laplacian of the diagram. For real kinematics, it is then customary to use the rescaling invariance $\alpha_e \to \lambda \alpha_e$ to integrate out the overall scale $\lambda$, which gives up to normalization
\be\label{eq:parametric}
\int_0^{1} \frac{\d^{\E}\alpha_e\, \delta(\sum_e \!\alpha_e - 1)}{\U^{\D/2}\, \V^{\E - \L\D/2}},
\ee
where $\E - \L \D/2$ is the superficial degree of divergence. It is understood that causality and convergence are imposed either by shifting $\V \to \V + i\eps$ with infinitesimal $\eps$, or as a contour deformation, see App.~\ref{app:contour}. Here we will \emph{not} deform $\alpha_e$'s, but instead implement causality by deforming the external kinematics such that $\Im \V > 0$ when a physical limit is approached. We complexify kinematics only after the representation \eqref{eq:parametric} is obtained. 

We will assume that the Feynman integral is free of overall divergences, which can be dealt with (for example in dimensional regularization) without affecting the singularities of the integral. We have to consider only diagrams dependent on at least one kinematic invariant, since others are either finite or excluded by the above assumption. Without loss of generality we consider one-particle irreducible diagrams from now on, since inclusion of reducible diagrams can at most introduce simple poles that do not affect the analytic continuation described below.

All quantities in \eqref{eq:parametric} can be expressed in terms of combinatorics of the Feynman diagram, see, e.g., \cite{nakanishi1971graph}. The action can be written as (sums in this letter always range in $i,j=1,2,\ldots,n$ and $e=1,2,\ldots,\E$):
\be\label{eq:V}
\V := - \sum_{i<j} p_i {\cdot} p_j\, {\cal G}_{ij} - \sum_e m_e^2\, \alpha_e,
\ee
where $m_e$ is the mass associated to the edge $e$ and $\G_{ij}$ is the Green's function (for a scalar field on a graph) between vertices where $p_i^\mu$ and $p_j^\mu$ enter the Feynman diagram. The graph Green's function $\G_{ij}$ measures the response of the diagram to changes in the dot product $p_i {\cdot} p_j$, and does not depend on $n$ but only on the topology of the diagram. This representation makes it obvious that we are dealing with a superposition of multiple two-point function problems, all jumbled up due to the momentum conservation \footnote{The representation \eqref{eq:V} makes it transparent why proving crossing symmetry off-shell is trivial in perturbation theory: all the $n(n{-}1)/2$ kinematic invariants can be deformed independently in $\Im p_i {\cdot} p_j < 0$ for $i \neq j$ while preserving analyticity and causality (see also \cite{deLacroix:2018tml} for the loop momentum perspective). It is imposing on-shellness, i.e., fixed $p_i^2 = M_i^2$, that introduces difficulties.}.

For completeness we give expressions for $\U$ and $\G_{ij}$, though they will not be needed directly in the proof. The polynomial $\U$ is given by
\be\label{eq:U}
\U := \sum_{T} \prod_{e \notin T} \alpha_e,
\ee
where the sum is over all spanning trees $T$ obtained by removing exactly $\L$ edges from the diagram.
The individual $\G_{ij}$'s can be written as
\be\label{eq:Gij}
\G_{ij} := \frac{1}{\U} \sum_{F_{ij}} \prod_{e \notin F_{ij}} \alpha_e,
\ee
which sums over all spanning two-forests $F_{ij}= T_i \sqcup T_j$, obtained by cutting $\L{+}1$ edges such that vertices where $p_i^\mu$ and $p_j^\mu$ enter the diagram belong to (possibly empty) trees $T_i$ and $T_j$ respectively. An example is given in App.~\ref{app:example}.

\subsection{Landau Equations}

Singularities of Feynman integrals are governed by Landau equations \cite{Landau:1959fi}, which in the representation \eqref{eq:parametric} read \cite{10.1143/PTP.22.128,Bjorken:1959fd}:
\be
\alpha_e \frac{\partial \V}{ \partial \alpha_e} = 0
\ee
for all edges $e$. Since a solution involving $\alpha_{e'} = 0$ gives Landau equations for a simpler graph with the edge $e'$ contracted, and we already take into account all graph topologies, we only need to consider \emph{leading} Landau equations with $\alpha_e \neq 0$ (in other words, the analytic continuation we will employ avoids sub-leading Landau singularities just as well as the leading ones). One can interpret them as the classical limit of the action $\V$ where all propagators go on-shell according to \eqref{eq:intro-V}. See App.~\ref{app:interpretation} for more details \footnote{In going from \eqref{eq:exponential} to \eqref{eq:parametric} we broke projective invariance, which obscures potential singularities at infinities \cite{doi:10.1063/1.1724262}. They need not concern us because we will not deform the integration contour.}. Recent work on Landau equations includes \cite{Chin:2018puw,Prlina:2018ukf,Collins:2020euz,Komatsu:2020sag,Muhlbauer:2020kut}.

Given the definition in \eqref{eq:V}, $\V$ is a degree-one homogeneous function in $\alpha_e$'s, which means on the solution of Landau equations we have
\be
\V = \sum_e \alpha_e \frac{\partial \V}{\partial \alpha_e} = 0.
\ee
This is a necessary (but not sufficient) condition for a singularity. Since leading Landau equations require $\alpha_e > 0$, the definitions \eqref{eq:U} and $\eqref{eq:Gij}$ give $\U>0$ and $\G_{ij}>0$.

\subsection{Upper Bound on the Graph Green's Functions}

\begin{figure}[!h]
	\includegraphics[scale=1.2]{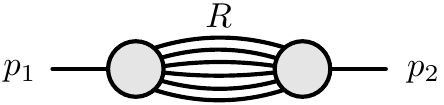}
	\caption{\label{fig:threshold}Normal threshold for $n=2$, where a subset $R$ of propagators goes on-shell.}
\end{figure}

In the following steps we will need an upper bound on $\G_{ij}$ that does not depend on the number of loops, edges, or external states.
As a proxy for its derivation, let us briefly consider the case $n=2$ off-shell, where $-p_1 {\cdot} p_2 = p_1^2$ is allowed to vary, and anomalous thresholds are absent. We have
\be
\V \,=\, p_1^2 \G_{12} - \sum_e m_e^2 \alpha_e \,\leqslant\, p_1^2 \G_{12} - m^2,
\ee
where $m$ is the lightest of $m_e>0$ and we used $\sum_e \alpha_e = 1$. Since $\V < 0$ for $p_1^2=0$, the action has to stay negative before encountering the first physical threshold at
\be
p_1^2 \,=\, \Big(\sum_{e\in R} m_e\Big)^{\!2} \,\geqslant\, |R|^2 m^2,
\ee
where $R$ is the set of $|R|$ intermediate particles, as in Fig.~\ref{fig:threshold}. This implies $\G_{12} \leqslant 1/|R|^2$. Since the labeling of the momenta was arbitrary, we have
\be\label{eq:Gij-bound}
\G_{ij} \leqslant \frac{1}{4},
\ee
because $|R| \geqslant 2$ for one-particle irreducible diagrams.
An alternative derivation using only graph theory is given in \cite{10.1143/PTPS.18.1}, which shows it holds without any restriction on masses.

We remind the reader that analyticity of higher-point on-shell amplitudes is not well-understood because of the presence of anomalous thresholds. The above trick circumvents this issue by deriving bounds on the individual $\G_{ij}$ which are the building blocks entering \eqref{eq:V} for arbitrary $n$.

\section{Bounds on Crossing Symmetry}

We will show how to analytically continue between two non-singular points in the real kinematic space, denoted by $p_i^{\zero} \!{\cdot} p_j^{\zero}$ and $p_i^{\one} \!{\cdot} p_j^{\one}$, i.e., where Landau equations are not satisfied. The deformation takes place in the $n(n{-}3)/2$-dimensional space of independent kinematic invariants $p_i {\cdot} p_j$, whose pre-image in the momentum vectors $p_i^\mu$ can be only realized in an $(n{-}1)$-dimensional space in the intermediate steps.

We preface the discussion with a naive approach in order to highlight why it is causality that puts bounds on crossing symmetry.

\subsection{Naive Approach}

\begin{figure}[!h]
	\includegraphics[scale=1.2]{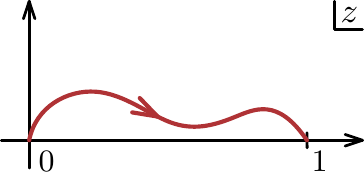}
	\caption{\label{fig:path}Path of deformation in the upper-half plane.}
\end{figure}

We introduce a complex variable $z$ and linearly deform the kinematic invariants according to
\be
p_i {\cdot} p_j = p_i^{\zero} \!{\cdot} p_j^{\zero} + z\, \big(p_i^{\one} \!{\cdot} p_j^{\one} - p_i^{\zero} \!{\cdot} p_j^{\zero}\big),
\ee
as well as consider a path in the upper-half plane approaching the two kinematic points at $z=0$ and $z=1$, see Fig.~\ref{fig:path}.
This deformation preserves momentum conservation and on-shell conditions, $p_i^2 = M_i^2$.
\vspace{-.15em}

Since $\V$ responds linearly to changes in kinematics, we have
\be
\V = \V_0 + z\, (\V_1 - \V_0),
\ee
where $\V_{\alpha} := \V(p_i^{\talpha} \!{\cdot} p_j^{\talpha})$. Remaining on the original integration contour with $\alpha_e$ real, the real and imaginary parts of the leading Landau equations are
\begin{gather}
\frac{\partial \V_0}{\partial \alpha_e} + \Re z \left( \frac{\partial \V_1}{\partial \alpha_e} - \frac{\partial \V_0}{\partial \alpha_e} \right) = 0,\\
\Im z \left( \frac{\partial \V_1}{\partial \alpha_e} - \frac{\partial \V_0}{\partial \alpha_e} \right) = 0,
\end{gather}
for all edges $e$. These cannot be \emph{simultaneously} satisfied along the deformation path with $\Im z > 0$: vanishing of the imaginary part implies $\partial \V_0 / \partial \alpha_e = 0$ for all $e$, which is a contradiction. Hence there are no singularities in the upper-half plane of $z$.

Nevertheless, this deformation cannot be used because by utilizing the imaginary part of $\V$ for the deformation we lost a reliable way of imposing the $i\eps$ prescription near both of two physical points. Put differently, the path of analytic continuation will in general veer away from the physical sheet.

An exception to this point are planar diagrams, which have vastly simpler analyticity properties and crossing for $n \leqslant \D{+}1$ can be proven without any constraints on masses, see App.~\ref{app:planar}.

\subsection{Fixing the $i\eps$}

In order to guarantee the correct $i\eps$ prescription we will add an intermediate step in the deformation, which passes through an open set $\{p_i^{\tast} \!{\cdot} p_j^{\tast}\}$ for which
\be
\V_{\ast} < 0
\ee
across the while integration contour. For the time being let us assume such points exist and deform
\be
p_i {\cdot} p_j = p_i^{\tast} \!{\cdot} p_j^{\tast} + z\, \big(p_i^{\one} \!{\cdot} p_j^{\one} - p_i^{\tast} \!{\cdot} p_j^{\tast}\big)
\ee
followed by an analogous continuation connecting $p_i^{\tast} \!{\cdot} p_j^{\tast}$ to $p_i^{\zero} \!{\cdot} p_j^{\zero}$. We further restrict to $\Re z \geqslant 0$. There are two cases depending on the sign of $\V_1 - \V_{\ast}$. When $\V_1 > \V_\ast$ we have
\be
\Im \V = \Im z\; (\V_1 - \V_{\ast}) > 0,
\ee
which is the correct causal prescription.  Otherwise, when $\V_1 \leqslant \V_\ast$, we have
\be
\Re \V = \V_\ast + \Re z\; (\V_1 - \V_{\ast}) < 0,
\ee
since $\V_{\ast} < 0$ by assumption. In those cases there is no singularity on the real axis and hence the $i\eps$ is not needed.

It remains to prove that a set of $\{p_i^{\tast} \!{\cdot} p_j^{\tast} \}$ with $\V_\ast < 0$ exists in the first place.

\subsection{Bounds on Masses}

We consider kinematics with $-p_i^{\tast} \!{\cdot} p_j^{\tast} < c$ for each of the $n(n{-}1)/2$ kinematic invariants appearing in $\V_{\ast}$ and some positive constant $c$. Using the upper bound on $\G_{ij}$ from \eqref{eq:Gij-bound} and calling $m=\min_e(m_e)$ the lightest internal mass one finds
\be
\V_\ast \,<\, \frac{n(n{-}1)}{8} c - m^2.
\ee
Therefore, requiring that $\V_{\ast} < 0$ yields
\be
-p_i^{\tast} \!{\cdot} p_j^{\tast} \,<\, c \,<\, \frac{8}{n(n{-}1)} m^2.
\ee
In terms of the external masses $M_i$ this translates to two types of constraints. Using momentum conservation requires on the one hand
\be
M_i^2 \,=\, -p_i^{\tast} {\cdot}\! \sum_{j \neq i} p_j^{\tast} \,<\, \frac{8}{n} m^2
\ee
and on the other
\be\label{eq:constraint}
\sum_{j} M_j^2 - 2M_i^2 \,=\, - \sum_{j \neq i} p_j^{\tast} {\cdot}\! \sum_{k \neq i,j} p_k^{\tast} \,<\, \frac{8(n{-}2)}{n} m^2
\ee
for all $i$. They together imply the constraints \eqref{eq:intro-m-constraint}.
These are the conditions for crossing symmetry to be satisfied to all loops and multiplicities.

\section{Outlook}

Let us comment on two natural directions for future work: optimizing the bounds and preventing the momenta $p_i^\mu$ from wandering into higher dimensions.

With respect to the former, let us notice that the upper bounds $\G_{ij} \leqslant 1/|R|^2$ are saturated on configurations where $\alpha_e \approx 1/|R|$ for each of the intermediate lines $e \in R$, and $\alpha_{e} \approx 0$ otherwise, cf. Fig.~\ref{fig:threshold}. Clearly, such bounds cannot be attained for all $\G_{ij}$'s simultaneously, because $\G_{ij}$ are not mutually independent (for example, they satisfy $\G_{ij} + \G_{jk} \geqslant \G_{ik}$). It is not unlikely that exploiting such inter-dependencies can improve bounds on crossing symmetry, though probably not significantly so for generic quantum field theories.
On the other hand, implementing conservation laws for specific processes might improve the bounds, perhaps along the lines of previous work on dispersion relations \cite{BOYLING1963249,BOYLING1964435}.

Remaining in four dimensions for $\Im z>0$ requires imposing vanishing of every $5{\times}5$ minor of $p_i {\cdot} p_j$ treated as a matrix, which would violate linear dependence on $z$ that our arguments hinged upon. Instead, one should employ a deformation directly on the four-momenta $p_i^\mu$ that correspond to linear shifts of $p_i {\cdot} p_j$, such as those used in on-shell recursion relations \cite{Britto:2005fq}. Nonetheless, in some situations it might be possible to get away without doing so, such as in the case of four-point scattering in two dimensions with equal external masses, $M_i = M$. In this setup we have $(p_1 {+} p_3)^2=0$ and repeating the steps from previous sections gives
\be
\V_\ast \,\leqslant\, \frac{1}{2}(s_{\ast}+M^2) - m^2
\ee
with $s_\ast = (p_1^{\tast}{+}p_2^{\tast})^2$.
Since we can choose $s_{\ast}$ to be arbitrarily small, it guarantees crossing symmetry and maximal analyticity for $m_e > M/\sqrt{2}$ to all loops.

\begin{acknowledgments}
The author thanks Edward Witten for illuminating discussions.
He gratefully acknowledges the funding provided by Frank and Peggy Taplin as well as the grant DE-SC0009988 from the U.S. Department of Energy.
\end{acknowledgments}

\section*{Appendices}
\appendix

\section{\label{app:conjecture}Why Crossing Symmetry is Still a Conjecture}

Here we give a lightning review of the core arguments of Gell-Mann, Goldberger, and Thirring \cite{GellMann:1954db} in a more modern formulation which can be found in \cite{Sommer:1970mr,Itzykson:1980rh}. We start with a local quantum field theory with a mass gap and consider (scalar for simplicity) charged fields $\varphi_a(-x)$ and $\varphi_b^\dagger(x)$ at spacelike separation, $x^2 < 0$. Following the Lehmann--Symanzik--Zimmerman procedure we introduce the currents
\begin{align}
	j_a(-x) &= (\Box_{-x} - M_a^2)\, {\cal \varphi}_{a}(-x),\\
	j_b^\dagger(x) &= (\Box_x - M_b^2)\, {\cal \varphi}_{b}^\dagger(x),
\end{align}
but do not take the on-shell limit until the very end. For the purposes of this discussion we will ignore irrelevant normalization factors.
The quantity of our interest is
\be\label{eq:quantity}
{\cal C} = \int \d^{4} x\,  e^{i (p_b - p_a) \cdot x} \langle \text{out}|
[j_b^\dagger(x),\, j_a(-x)]
| \text{in} \rangle
\ee
for $p_a^\mu$ and $p_b^\mu$ future and past timelike respectively. The remaining $n{-}2$ states $\langle\text{out}|$ and $|\text{in}\rangle$ are arbitrary.
We can expand the commutator in two different ways. One of them separates $x^\mu$ into the future and past lightcone,
\be\label{eq:two-terms}
[j_b^\dagger(x),\, j_a(-x)] = \theta(x^0) [j_b^\dagger(x),\, j_a(-x)] + \theta(-x^0) [j_b^\dagger(x),\, j_a(-x)].
\ee

The first term is a retarded commutator, which in the integrand can be replaced by a time-ordered product of the two currents, giving
\be\label{eq:first-crossed}
\int \text{d}^{4} x\,  e^{i (p_b - p_a) \cdot x} \langle \text{out}|
{\cal T} j_b^\dagger(x)\, j_a(-x)
| \text{in} \rangle.
\ee
It is the Green's function for the process $\{ \mathrm{in}, p_a \} \to \{\mathrm{out}, -p_b\}$ with the overall momentum conservation delta function stripped away. Assuming causality and temperedness (polynomial boundedness) of the bra-ket, the integrand has support only when $x^\mu$ is a future timelike vector. This implies that the Fourier transform is convergent when $\Im (p_b {-} p_a)$ is future timelike, since only then the integrand is damped by a factor of $e^{-\Im(p_b {-}p_a)\cdot x}$.

On the other hand, after relabeling $x \to -x$, the second term in \eqref{eq:two-terms} is a retarded commutator for the crossed process, giving
\be\label{eq:second-crossed}
-\int \text{d}^{4} x\,  e^{i (p_a - p_b) \cdot x} \langle \text{out}|
{\cal T} j_a(x)\, j_b^\dagger(-x)
| \text{in} \rangle.
\ee
The additional minus sign came from reversing the commutator. This is the Green's function for $\{ \mathrm{in}, \overline{p}_b \} \to \{\mathrm{out}, -\overline{p}_a\}$ scattering, where the bar denotes an anti-particle. It suggests that $\cal C$ might be the difference between the two Green's functions. However, repeating previous arguments one finds that the integral \eqref{eq:second-crossed} converges only when $\Im (p_b {-} p_a)$ is \emph{past} timelike, which has no overlap with the domain of analyticity of \eqref{eq:first-crossed}. Therefore in order to relate the two processes one needs to show that $\cal C$ can be analytically continued between the two kinematic regions.

To this end, let us evaluate $\cal C$ using the more obvious way of expressing the commutator,
\be
[j_b^\dagger(x),\, j_a(-x)] = j_b^\dagger(x)\, j_a(-x) - j_a(-x)\, j_b^\dagger(x).
\ee
We then use unitarity to insert a complete basis of states $\mathds{1} = \int \d^{4} p_I \sum_I | I \rangle\langle I |$ and translation invariance so that $\cal C$ evaluates to
\begin{align}
&\sum_I \langle \text{out}| j_b^\dagger(0)| I \rangle \,\langle I | j_a(0) | \text{in} \rangle\, \delta^{4} (p_b {-} p_a {-}p_{\mathrm{in}} {+} p_{\mathrm{out}} {-} 2p_I) \nn\\
&-\sum_I \langle \text{out}| j_a(0)| I \rangle \,\langle I | j_b^\dagger(0) | \text{in} \rangle\, \delta^{4} (p_b {-} p_a {+}p_{\mathrm{in}} {-} p_{\mathrm{out}} {+} 2p_I),
\end{align}
where $p_I^\mu$ are the momenta of the intermediate states, which by assumption of the non-zero mass gap satisfy $p_I^2 >0$. Therefore ${\cal C} =0$ in a region $\Phi$ of the $\Re (p_b {-} p_a)$ space that lies below the production threshold for the lightest intermediate states in both sums.

Since the regions of analyticity of \eqref{eq:first-crossed} and \eqref{eq:second-crossed} border $\Phi$, within which ${\cal C}$ vanishes, the edge-of-the-wedge theorem  guarantees that the two crossed processes must be analytic continuations of each other in the region where $\Re (p_b {-} p_a) \in \Phi$ and $\Im (p_b {-} p_a)$ belongs to the union of future and past lightcones.

This argument does not yet imply crossing symmetry for two reasons. Firstly, the above domain of analyticity has no intersection with on-shell kinematics. In order to see this we can use a (complex) Lorentz frame, such that in lightcone coordinates
\be
p_a^\mu = (M_a, M_a, 0, 0), \qquad p_b^\mu = (p_b^+, p_b^-, 0, 0),
\ee
where $M_a > 0$.
The constraint of $\Im (p_b {-} p_a)$ being timelike implies
\be\label{eq:lightcone1}
(\Im p_b^+)(\Im p_b^-) > 0.
\ee
However, on-shell we need $p_b^+ p_b^- = M_b^2 > 0$, whose real and imaginary parts give respectively
\begin{gather}
(\Re p_b^+)(\Re p_b^-) - (\Im p_b^+)(\Im p_b^-) > 0,\label{eq:lightcone2}\\
(\Re p_b^+)(\Im p_b^-) + (\Im p_b^+)(\Re p_b^-) = 0.\label{eq:lightcone3}
\end{gather}
The constraint \eqref{eq:lightcone1} together with \eqref{eq:lightcone3} mean that $\Re p_b^+$ and $\Re p_b^-$ have to have opposite signs, which is in contradiction with \eqref{eq:lightcone2}.

The second problem concerns the assumption of temperedness made before, which was not necessarily justified. To check it, one ought to consider the setup where $\langle \mathrm{in} |$ and $| \mathrm{out} \rangle$ are vacuum states and all $n$ particles are represented as fields. Repeating essentially the same steps as above with more book-keeping, one arrives at the so-called \emph{primitive domain} of analyticity (with no support on-shell) \cite{Steinmann1960a,Steinmann1960b,ruelle1961connection,doi:10.1063/1.1703695,araki1960properties}, where all crossed Green's functions agree. It is then a geometric problem to show that the envelope of holomorphy of the primitive region intersects the real on-shell regions for all crossed processes \cite{Bros:1964iho,Bros:1972jh}. Moreover, for $2 \to 2$ and $2 \to 3$ processes specifically, it has been shown that physical regions are connected via asymptotic domains with large complex center-of-mass energy, which implies crossing symmetry for on-shell scattering amplitudes \cite{Bros:1965kbd,Bros:1985gy}.

\section{\label{app:contour}Causal Contour Deformations}

Let us review how to implement infinitesimal contour deformation that replaces the $i\eps$ prescription. We start by deforming Schwinger parameters $\alpha_e$ into
\be
\hat{\alpha}_e = \frac{\alpha_e + i \eps \beta_e}{1 + i \eps \sum_{e'} \beta_{e'}},
\ee
which satisfy $\sum_e \hat{\alpha}_e = 1$. Here $\beta_e = \beta_e(\alpha_{e'})$ are yet to be determined functions, which are assumed to be $\eps$-independent and vanish at $\alpha_e=0$ and $\alpha_e=1$ in order not to alter the endpoints of integration. Using homogeneity of $\V$, the deformed $\hat\V = \V(\hat\alpha_e)$ reads
\be
	\hat\V = \frac{\V(\alpha_e + i \eps \beta_e)}{1+ i \eps \sum_{e} \beta_{e}}.
\ee
Expanding in $\eps$ gives
\begin{align}
\hat{\V} &= (1- i \eps {\textstyle\sum}_{e} \beta_{e}) \left(\V + i\eps \sum_{e'} \beta_{e'} \frac{\partial \V}{\partial \alpha_{e'}} \right)  + {\cal O}(\eps^{2})\nn\\
	&= \V + i\eps \sum_{e} \beta_{e} \left(\frac{\partial \V}{\partial \alpha_{e}} - \V \right) + {\cal O}(\eps^{2}).
\end{align}
This suggest a natural choice for $\beta_e$,
\be
\beta_e = \alpha_e (1{-}\alpha_e) \left(\frac{\partial \V}{\partial \alpha_{e}} - \V \right),
\ee
which satisfies all the required properties and gives the deformed action
\be
\hat{\V} = \V + i\eps \sum_{e} \alpha_e (1{-}\alpha_e) \left(\frac{\partial \V}{\partial \alpha_{e}} - \V \right)^{\!2} + {\cal O}(\eps^{2}).
\ee
Hence for sufficiently small $\eps$ and real kinematics, it implements $\Im \hat{\V}>0$ unless Landau equations are satisfied. Similar deformations are routinely used in numerical evaluations of Feynman integrals, see, e.g., \cite{Nagy:2006xy}.

\section{\label{app:example}Acnode Diagram Example}

\begin{figure}[!h]
	\includegraphics[scale=1.2]{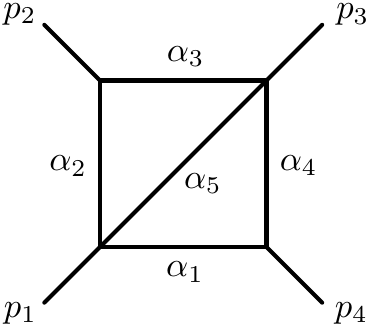}
	\caption{\label{fig:acnode}Acnode diagram.}
\end{figure}

Let us consider a classic example of the acnode diagram, illustrated in Fig.~\ref{fig:acnode}.
According to the definition \eqref{eq:U}, $\U$ is given as a sum over eight spanning trees,
\be
\U = \alpha_5(\alpha_1 {+} \alpha_2 {+} \alpha_3 {+} \alpha_4) + (\alpha_2 {+} \alpha_3)(\alpha_4 {+} \alpha_1).
\ee
The graph Green's functions $\G_{ij}$ between vertices where $p_i^\mu$ and $p_j^\mu$ enter the diagram are given by the expressions \eqref{eq:Gij}, which yield explicitly
\begin{gather}
\G_{12} = \tfrac{1}{\U} \alpha_2 [(\alpha_1 {+} \alpha_4)(\alpha_3{+}\alpha_5) + \alpha_3 \alpha_5], \\
\G_{13} = \tfrac{1}{\U} \alpha_5(\alpha_1 {+} \alpha_4)(\alpha_2 {+} \alpha_3), \\
\G_{14} = \tfrac{1}{\U}\alpha_1 [(\alpha_2 {+} \alpha_3)(\alpha_4 {+} \alpha_5) + \alpha_4 \alpha_5],\\
\G_{23} = \tfrac{1}{\U} \alpha_3 [(\alpha_1 {+} \alpha_4)(\alpha_2 {+} \alpha_5) + \alpha_2 \alpha_5],\\
\G_{24} = \tfrac{1}{\U}[(\alpha_1{+}\alpha_2)(\alpha_3 \alpha_4 {+} \alpha_3 \alpha_5 {+} \alpha_4 \alpha_5) + \alpha_1 \alpha_2 (\alpha_3 {+} \alpha_4)],\\
\G_{34} = \tfrac{1}{\U} \alpha_4 [(\alpha_1 {+} \alpha_5)(\alpha_2 {+} \alpha_3) + \alpha_1 \alpha_5].
\end{gather}
One can confirm that on the support of the constraint $\sum_e \alpha_e = 1$, all the $\G_{ij}$'s are upper-bounded by $1/4$, except for $\G_{13}$, whose bound can be improved to $1/9$ due to the topology of the diagram.

The acnode diagram owes its name to a type of singularity that appears on the physical sheet for a range of masses \cite{doi:10.1063/1.1703752,doi:10.1063/1.1664557}. It was previously studied for $M_1 = M_3 = M$ and all the remaining $M_i = m_e = m$. The acnodes and real cusps appear when
\be
m \leqslant \frac{M}{\sqrt{4{+}2\sqrt{2}}} \simeq 0.383 M
\ee
and hence below our bound \eqref{eq:intro-bound-45}.

\section{\label{app:interpretation}Physical Interpretation of Landau Equations}

Landau equations can be interpreted as the classical limit of Feynman integrals, in which on-shell propagators describe particles traveling in space-time \cite{Coleman:1965xm,Nakanishi:1968hu}. For the purpose of the discussion below we specialize to the strictly massive case, $m_e >0$.

One can show that integrating out loop momenta leads to the following constraints on $q_e^\mu$ in terms of Schwinger parameters $\alpha_{e'}$. Let us assign arbitrary orientations to each internal edge. Momentum conservation at each of the $\mathrm{V}$ vertices $v$ reads
\be\label{eq:momentum-conservation}
p_v^\mu + \sum_{e} \eta_{ve}\, q_e^\mu = 0,
\ee
where $\eta_{ve}$ equals $+1$ when $q_e^\mu$ is incoming towards $v$, $-1$ when it is outgoing, and $0$ otherwise. Here $p_v^\mu$ is the total external momentum flowing into the vertex $v$. Similarly, we have the conservation law for each of the $\mathrm{L}$ oriented loops $\ell$,
\be\label{eq:conservation}
\sum_e \alpha_e\, \eta_{\ell e}\, q_e^\mu = 0,
\ee
where $\eta_{\ell e}$ equals $+1$ when the orientations of the loop $\ell$ and the edge $e$ agree, $-1$ when they disagree, and $0$ when the edge does not belong to the loop. After solving $q_e^\mu = q_e^\mu(\alpha_{e'}\!)$ in terms of Schwinger parameters and external kinematics we have
\be
\V = \sum_e \alpha_e (q_e^2 - m_e^2).
\ee
Since $q_e^2$ are degree-zero homogeneous functions of $\alpha_{e'}\!$'s, the leading Landau equations are equivalent to putting all propagators on-shell,
\be\label{eq:on-shell}
q_e^2 - m_e^2 = 0.
\ee

Let us trivialize the constraints \eqref{eq:conservation} by introducing a Lorentz vector $x_v^\mu$ associated to each vertex $v$. Calling $\Delta x_e^\mu$ the difference between $x_v$'s at the end and beginning of the edge $e$ we assign
\be
\alpha_e q_e^\mu = \tfrac{1}{2}\Delta x_e^\mu = \tfrac{1}{2}\sum_v \eta_{ve}\, x_v^\mu,
\ee
which automatically satisfies \eqref{eq:conservation}. In the classical limit each $x_v^\mu$ has an interpretation of position of the vertex $v$ in space-time. Let us confirm this by evaluating a scalar Feynman integral in position space. Up to normalization it can be written as
\be\label{eq:position-space}
\int \d^\mathrm{\D V} x_v\, e^{-i \sum_v p_v \cdot x_v / \hbar} \prod_{e} \mathrm{G}_F (\Delta x_e, m_e) ,
\ee
where $\mathrm{G}_F(\Delta x_e,m_e)$ denotes the Feynman propagator between points at timelike separation $\Delta x_e$ and mass $m_e$,
\be
\mathrm{G}_F(\Delta x_e,m_e) = \int \frac{\d^\D q_e}{(2\pi)^\D} \frac{e^{-i q_e \cdot \Delta x_e / \hbar}}{q_e^2 - m_e^2 + i\eps} .
\ee
It can be expressed as a Bessel function, which up to overall normalization reads
\be
\int_{0}^{\infty} \frac{\d \alpha_{e}}{\alpha_e^{\D/2}} \exp \left[ - \frac{i}{\hbar}\left(  \frac{\Delta x_{e}^2}{4\alpha_e} + \alpha_e(m_e^2 {-} i\eps)\right)\right].
\ee
Therefore the Feynman integral in \eqref{eq:position-space} evaluates to
\be
\int \frac{\d^\mathrm{\D V} x_v\, \d^\mathrm{E} \alpha_e}{(\prod_e\! \alpha_e)^{\D/2}}\, e^{i\V /\hbar},
\ee
where, ignoring the $i\eps$ factor, $\V$ in the exponent is given by
\be\label{eq:V-appendix}
\V = - \sum_v p_v {\cdot} x_v - \sum_e \left( \frac{\Delta x_e^2}{4\alpha_e} + m_e^2 \alpha_e \right).
\ee
Extremizing $\V$ with respect to $x_v$ yields
\be
p_v^\mu + \sum_e \eta_{ve} \frac{\Delta x_e^\mu}{2\alpha_e} = 0
\ee
for each vertex $v$. This is just the momentum conservation \eqref{eq:momentum-conservation}. Similarly, varying $\alpha_e$ we get
\be\label{eq:proper-time}
\frac{\Delta x_e^2}{4\alpha_e^2} - m_e^2 = 0
\ee
for all edges $e$, which are the on-shell conditions \eqref{eq:on-shell}. We conclude that the classical limit in the position space reproduces Landau equations and $x_v^\mu$ can be interpreted as positions where particles interact in space-time. This is the origin of the term ``Schwinger proper time'' since according to \eqref{eq:proper-time} the $\alpha_e$'s measure the proper time $\sqrt{\Delta x_e^2}$ elapsed between two interactions, normalized by the particle mass $m_e$.

While the most obvious saddles of \eqref{eq:V-appendix} are those lying on the original integration contour with $\alpha_e$ and $x_v^\mu$ real, in general solutions of Landau equations are complex. Their physical meaning remains nebulous. See \cite{Passarino:2018wix} for a recent review of observable effects of anomalous thresholds at particle colliders.

For completeness let us mention that Landau equations have an interpretation in terms of electric circuits \cite{Mathews:1959zz,Bjorken:1959fd}. This should not be surprising because both can be described as a scalar field on a graph. In this interpretation, component by component, $x_v^\mu$ measures voltage at $v$, $q_e^\mu$ is the current flowing through $e$, and $\alpha_e$ its resistance. Then \eqref{eq:momentum-conservation} and \eqref{eq:conservation} are the Kirchhoff's circuit laws. When a unit current is applied flowing from the vertex where $p_i^\mu$ enters the diagram to the one where $p_j^\mu$ does so, the graph Green's function
\be
\G_{ij} = \sum_e \alpha_e q_e^2
\ee
measures the power dissipated in the circuit.

\section{\label{app:planar}Crossing Symmetry for Planar Diagrams}

An alternative representation of $\V$ which is particularly suitable for planar diagrams reads
\be\label{eq:V-alt}
\V \,=\, \sum_S p_S^2\, \F_S - \sum_e m_e^2\, \alpha_e,
\ee
where the first sum goes over all $2^{n-1} {-} 1$ proper subsets of external particles $S$, without double-counting the complements $\bar{S} = \{1,2,\ldots,n\} {\setminus} S$. Here
\be\label{eq:Fs}
\F_S \,=\, \F_{\bar{S}} \,:= \, \frac{1}{\U} \sum_{F_S} \prod_{e \notin F_S} \alpha_e
\ee
involve sums over all two-forests $F_S = T_S \sqcup T_{\bar{S}}$ such that $T_S$ and $T_{\bar{S}}$ only contain vertices where particles from the sets $S$ and $\bar{S}$ enter the diagram respectively. For example, for the diagram in Fig.~\ref{fig:acnode} we have
\begin{gather}
	\F_{1} = \tfrac{1}{\U}\alpha_1 \alpha_2 \alpha_5, \quad \F_{2} = \tfrac{1}{\U} \alpha_2 \alpha_3 (\alpha_1 {+} \alpha_4 {+} \alpha_5), \\
	\F_{3} = \tfrac{1}{\U} \alpha_3 \alpha_4 \alpha_5, \quad \F_{4} = \tfrac{1}{\U} \alpha_1 \alpha_4 (\alpha_2 {+} \alpha_3 {+} \alpha_5),\\
	\F_{12} = \tfrac{1}{\U} \alpha_1 \alpha_3 \alpha_5,\quad \F_{23} = \tfrac{1}{\U} \alpha_2 \alpha_4 \alpha_5,\quad \F_{13} = 0,
\end{gather}
where $\F_{13}$ vanishes due to planarity of the diagram.
Comparing the definitions \eqref{eq:Gij} and \eqref{eq:Fs} it is easily seen that $\G_{ij}$ can be expressed as sums
\be
\G_{ij} = \sum_{S \not\ni \{i,j\} } \F_{S \cup i},
\ee
which range over all $2^{n-2}$ sets $S$ not including labels $i$ and $j$.

What is special about planar amplitudes is that, not counting $p_i^2 = M_i^2$, only $n(n{-}3)/2$ planar invariants $p_S^2$ appear in \eqref{eq:V-alt}, where $S$ have the appropriate planar ordering. It means they can be deformed independently of each other in the upper-half planes $\Im p_S^2 > 0$ (by previous assumptions the amplitude depends on at least one kinematic invariant). This argument requires that $p_i^\mu$ are embedded in at least $n{-}1$ dimensions, so that no additional constraints are put on the kinematics $p_i {\cdot} p_j$. Since $\F_S > 0$ for each $S$, along such a deformation we have
\be
\Im \V > 0,
\ee
which at the same time imposes the correct $i\eps$ prescription and allows one to deform between any two points in the real kinematic space, thus proving crossing symmetry without any constraints on masses.

In the case of the acnode diagram in $\D \geqslant 3$ the above prescription deforms the kinematics $s = (p_1 {+} p_2)^2$, $t=(p_2 {+} p_3)^2$ in $\Im s / \Im t >0$, while all the non-analyticity (beyond a certain mass threshold) are confined to $\Im s / \Im t < 0$ \cite{doi:10.1063/1.1703752}.

\bibliography{references}

\end{document}